\documentclass[12pt]{article}
\usepackage[nosort]{cite}
 \usepackage[nottoc]{tocbibind}
\usepackage[colorlinks=true,linkcolor=blue,urlcolor=magenta, citecolor=blue]{hyperref}
\usepackage{fullpage}
\usepackage{amssymb,graphicx}
\usepackage{amsmath}
\usepackage[margin=0.5cm]{caption}
\usepackage{hyperref} 
\usepackage{upgreek}

\def\d{\partial}
\def\l{\left(}
\def\r{\right)}

\newcommand{\be}{\begin{equation}}
\newcommand{\ee}{\end{equation}}
\newcommand{\bea}{\begin{eqnarray}}
\newcommand{\eea}{\end{eqnarray}}
\newcommand{\bg}{\begin{gather}}
\newcommand{\eg}{\end{gather}}
\newcommand{\bseq}{\begin{subequations}}
\newcommand{\eseq}{\end{subequations}}

\newcommand{\Tr}{{\rm Tr}}

\begin{document}
\baselineskip=15.5pt
\begin{titlepage}
\begin{center}
{\Large\bf   From QCD Strings to WZW}\\
\vspace{0.5cm}
{ \large
John C. Donahue$^{a}$, Sergei Dubovsky$^{a}$,  Guzm\'an Hern\'andez-Chifflet$^{a,b}$,\\
\vspace{.2cm} and Sergey Monin$^c$
}\\
\vspace{.45cm}
{\small  \textit{  $^a$Center for Cosmology and Particle Physics,\\ Department of Physics,
      New York University\\
      New York, NY, 10003, USA}}\\ 
      \vspace{.1cm}
      \vspace{.25cm}       
      {\small  \textit{   $^b$Instituto de F\'isica, Facultad de Ingenier\'ia,\\ Universidad de la Rep\'ublica,\\
      Montevideo, 11300, Uruguay}}\\ 
       \vspace{.1cm}
      \vspace{.25cm}       
      {\small  \textit{   $^c$William I. Fine Theoretical Physics Institute, \\University of Minnesota, Minneapolis, MN 55455, USA}}\\
\end{center}
\begin{center}
\begin{abstract}
According to the Axionic String Anstaz (ASA) confining flux tubes in pure gluodynamics 
 are in the same equivalence class as a new family of integrable non-critical strings,  called axionic strings. In addition to translational modes, axionic strings
carry a set of worldsheet axions transforming as an antisymmetric tensor under the group of transverse rotations. We initiate a study
of integrable axionic strings at general number of space-time dimensions $D$. We show that in the infinite tension limit worldsheet axions should be described by a peculiar ``pseudofree" theory---their $S$-matrix is trivial, but the corresponding action cannot be brought into a free form by a local field redefinition. This requirement fixes the axionic action to take a form of the $O(D-2)$ Wess--Zumino--Witten (WZW) model.

\end{abstract}
\end{center}
\end{titlepage}
%\tableofcontents
\newpage
\section{Introduction and Summary}
Understanding confining strings stands out as one of a few remaining major open problems in the Standard Model of particle physics that can be
solved while waiting for a new collider to be built to shed light on more murky issues, such as the hierarchy problem. Thinking about this problem has proven to be extremely fruitful in the past.  Among other things this lead to  discoveries of critical string theory \cite{Veneziano:1968yb}, the large $N$ expansion
\cite{tHooft:1973alw}, the Polyakov action \cite{Polyakov:1981rd}, holography \cite{Maldacena:1997re,Gubser:1998bc,Witten:1998qj,Polyakov:1998ju} and integrability of $N=4$ supersymmetric Yang--Mills \cite{Minahan:2002ve}.

Over the last two decades holography has been considered the most promising approach for constructing a quantitative  description of confining strings.  Its major qualitative prediction---the existence of an additional worldsheet scalar mode, corresponding to the warped holographic direction---perfectly matches the expectation from the Liouville description of non-critical strings. 

However, there is mounting evidence that non-critical strings describing  confining flux tubes in non-supersymmetric gluodynamics  are different. By now a considerable amount of lattice data on the excitation spectrum of confining flux tubes in $D=3,4$-dimensional $SU(N)$ Yang--Mills theory has been accumulated \cite{Athenodorou:2010cs,Athenodorou:2011rx,Athenodorou:2013ioa,Athenodorou:2016kpd,Athenodorou:2017cmw}. This data does not show any sign of a scalar excitation on the worldsheet of confinings strings in the fundamental representation of the gauge group\footnote{Massive scalar breathing modes are present on the worldsheet of flux tubes in higher representations of the gauge group \cite{Dubovsky:2014fma} (``$k$-strings"). These are unrelated to the holographic direction and indicate that $k$-strings are bound states of fundamental flux tubes. }. Instead, a massive {\it pseudoscalar} mode has been found at $D=4$ \cite{Dubovsky:2013gi}, while $D=3$ data is consistent with a massless translational Goldstone being the only degree of freedom on the string worldsheet \cite{Dubovsky:2014fma}. The latter conclusion is strongly supported also by the analysis \cite{Dubovsky:2016cog} of the glueball spectrum in $D=3$ gluodynamics \cite{Athenodorou:2016ebg}. Finally,  the analytically tractable $D=2$ case also supports the conclusion that confining strings in QCD-like theories are different from what one may expect based on holography in the (super)gravity approximation\cite{Dubovsky:2018dlk}. 

Let us stress that this disagreement does not imply that the holographic intuition is  completely useless for QCD-like theories. Instead, more likely this is just another indication that the (yet to be found) string theory background holographically dual to the real world QCD is very strongly curved, so that the (super)gravity description is not adequate and the full power of string theory is required. In the meantime, holography does serve as a useful inspiration for phenomenological string models \cite{Sonnenschein:2018fph}. Still, new ideas are clearly needed to construct a non-critical string theory describing confining flux tubes.

A concrete proposal in this direction---the Axionic String Ansatz (ASA)---has been put forward in \cite{Dubovsky:2015zey,Dubovsky:2016cog}. It is based on the observation that both in $D=3$ and $D=4$ Yang--Mills the matter content on the worldsheet, as observed with the current lattice data, 
matches the one of an integrable theory enjoying target space Poincar\'e symmetry $ISO(1,D-1)$. In both cases the integrable phase shift between any two scattering particles is of the Dray--'t Hooft form
\cite{Dray:1984ha},
\be
\label{eis}
e^{2i\delta(s)}=e^{i\ell_s^2 s/4}\;,
\ee
where $1/\ell_s^{2}$ is the string tension.
At $D=3$ the corresponding integrable theory contains a single massless scalar boson $X$ (the Goldstone mode of the string). At $D=4$ there are two massless scalar 
Goldstones $X^1$, $X^2$ and a massless pseudoscalar axion. In both cases the integrability on the confining string worldsheet is not exact---at $D=4$ it is broken by the axion mass, and at $D=3$  deviations from the phase shift (\ref{eis}) \cite{Dubovsky:2014fma}  as well as non-vanishing multiparticle amplitudes \cite{Chen:2018keo} have been extracted from lattice data.

At first sight the idea of approximate integrability of confining strings may appear completely  ad hoc. However, approximate integrability has prominently emerged in the past in several  perturbative QCD contexts  \cite{Lipatov:1993yb,Faddeev:1994zg,Korchemsky:1994um,Braun:1999te,Gorsky:2002ju,Minahan:2002ve,Ferretti:2004ba,Beisert:2004fv}. In the worldsheet scattering, approximate integrability at low energies directly follows from the non-linearly realized target space Poincar\'e symmetry \cite{Dubovsky:2012sh,Dubovsky:2014fma}. A more surprising aspect of the ASA is that integrability is expected to get restored also at high energies\footnote{Here and in what follows the large $N$ limit is implied, which makes it possible to define the high energy, $E\gg \Lambda_{QCD}$, asymptotics of the worldsheet theory. For details see, e.g., \cite{Dubovsky:2015zey}. }. This can be understood \cite{Dubovsky:2018vde} by identifying high energy worldsheet excitations with partons of perturbative QCD. Asymptotic freedom implies that their hard scattering is trivial at high energies and the worldsheet scattering is dominated by linearly growing time delays, associated with the phase shift (\ref{eis}), and caused by long strings stretched between the partons.

Within the ASA approach the first step towards a quantitative understanding of confining strings is to build a comprehensive description of integrable axionic strings. For instance, these are expected to produce a spectrum of short strings (glueballs) with exact degeneracies at each level similarly to conventional critical strings. The actual glueball spectra exhibit a well pronounced level structure, but  level degeneracies are only approximate \cite{Dubovsky:2016cog}. After integrable axionic strings are well understood it should be possible to calculate the corresponding splittings using various perturbative approximations, such as the large $J$ expansion \cite{Hellerman:2013kba}.
With this program in mind, our goal here is to develop a better understanding of integrable axionic strings.

 To start with, it is instructive to compare, following \cite{Dubovsky:2015zey}, axionic strings to the conventional non-critical strings \cite{Polyakov:1981rd}. Integrability provides a natural language to describe both on the same ground. In this language conventional critical strings are defined by the integrable  $S$-matrix (\ref{eis}) \cite{Dubovsky:2012wk} describing scattering on a worldsheet of a single infinitely long string. In a non-critical case $D\neq 26$, and in the absence of additional massless excitations on the worldsheet, integrability is necessarily broken by the universal one-loop particle production \cite{Dubovsky:2012sh,Cooper:2014noa} associated with the Polchinski--Strominger  (PS) term \cite{Polchinski:1991ax}. The only 
notable exceptional case is $D=3$ strings, where the PS particle production vanishes as a result of kinematical cancelations.

It is natural  to then ask what kind of additional massless matter can be added on the worldsheet to allow for an integrable theory enjoying the non-linearly realized target space Poincar\'e symmetry $ISO(1,D-1)$. In principle, the number of options is quite large. For instance, one may add a compact $c=26-D$ CFT, which does not transform under $ISO(1,D-1)$. This corresponds to considering a conventional compactification of critical bosonic strings. Alternatively, one may add fermions which transform non-trivially under $ISO(1,D-1)$. Depending on the choice of the fermion representation, one may reproduce this way either the Ramond--Neveu--Schwarz (RNS) or Green--Schwarz description of critical $D=10$ superstrings\footnote{In the conventional formalism \cite{Polchinski:1998rr} the critical central charge $c=15$ in the RNS case  is different from the bosonic $c=26$ value because of the extended gauged (super)symmetry resulting in a different (super)ghost system. In the integrability language the difference can be traced to different    $ISO(1,D-1)$ transformation properties of additional massless excitations, c.f. \cite{Mohsen:2016lch}.}.

The ``old-fashioned" non-critical strings \cite{Polyakov:1981rd}  in a sense correspond to the minimal option available at any $D$---one introduces a single massless scalar field $\phi$ and makes use of the linear dilaton coupling $\int \phi R$ to cancel the PS particle production.

However, as noticed in  \cite{Dubovsky:2015zey}, at $D=4$ another equally minimal option is available. Namely, one introduces a single massless {\it pseudoscalar} field $a$
(the worldsheet axion)
and makes use of the coupling to the string self-intersection number \cite{Polyakov:1986cs} to cancel particle production,
\be
\label{Sa}
S_a=Q_a\int d^2\sigma\;a\epsilon_{ij}K^i_{\alpha\gamma}K^{j\gamma}_\beta\epsilon^{\alpha\beta}\;,
\ee
where $K^i_{\alpha\gamma}$ is the worldsheet extrinsic curvature. Intriguingly, the value of the coupling constant $Q_a$ required for integrability,
\be
\label{Qa}
Q_a=\sqrt{7\over 16\pi}\approx 0.373176\dots
\ee
agrees within error bars with the value of the corresponding coupling constant for the massive worldsheet axion, as extracted from the lattice data \cite{Dubovsky:2013gi},
\[
Q_{lattice}\approx0.38\pm 0.04\;.
\]
This piece of numerology provided the initial motivation for the ASA. Note that the worldsheet axion is a very natural degree of freedom in the context of $D=4$ confining strings \cite{Dubovsky:2018vde}---it is created by an insertion of a transverse plaquette into a Wilson line operator,
\[
{\cal O}_a=Pe^{\int_{-\infty}^\infty dz A_z}F_{xy}\;.
\]

It was suggested in \cite{Dubovsky:2015zey} that it is natural to think about $D=3$ and $D=4$ integrable axionic strings as members of a family of non-critical integrable strings  which can be defined for a general  $D$ in the following way.  In addition to translational Goldstones $X^i$  one introduces a set of worldsheet pseudoscalars $A^{ij}=-A^{ji}$ ($i,j=1,\dots, D-2$) transforming as an antisymmetric tensor under the $O(D-2)$ group of unbroken transverse rotations. The axionic coupling (\ref{Sa}) is replaced by
\be
\label{SA}
S_A=Q_A\int d^2\sigma A_{ij}K^i_{\alpha\gamma}K^{j\gamma}_\beta\epsilon^{\alpha\beta}\;.
\ee

The goal of the present paper is to initiate a detailed study of this proposal. Clearly if such a family of integrable models indeed existed it would be very interesting, independently of the expected connection to confining strings. In addition, this is likely to provide a better understanding of the physically relevant $D=3$ and $D=4$ cases.
Indeed, these cases on their own are quite degenerate---at $D=3$ an antisymmetric tensor does not carry any local degrees of freedom and at $D=4$ it is equivalent to a pseudoscalar.
Understanding a non-degenerate $D>4$ case is likely to provide a further insight in the structure of axionic strings.
 Furthermore, for many purposes it has been fruitful to consider a formal analytic continuation of quantum field theories in $D$. Hence, if the relation between axionic and confining strings is correct, one may expect the (formal) analytic continuation of axionic strings in $D$ to exist, mirroring the analytic continuation of  Yang-Mills theory.

A priori, it is not obvious that for integrable axionic strings at $D>4$ all the scattering should be described just by the phase shift (\ref{eis}). In particular, axion self-interactions 
may be different. However, in this paper by $D$-dimensional axionic strings  we will mean the theory where the whole $S$-matrix is given just by the universal diagonal phase shift (\ref{eis}). This is what happens in other integrable examples mentioned above\footnote{If the $S$-matrix can be defined at all, which  may be not the case for compactifications with an interacting internal CFT.}, and presents the simplest generalization of $D=3$ and $D=4$ models.

As we explain in section~\ref{sec:symmetry}, axionic strings at general $D$ are much more subtle than the $D=4$ ones already at the tree level. Indeed,  at the leading order in derivative expansion the action 
of $D=4$ axionic strings is simply the $D=5$   Nambu--Goto action with an axion $a$ entering as an additional coordinate. This is no longer compatible with a non-linearly realized $ISO(1,D-1)$ symmetry at general $D$ 
because axions $A^{ij}$ transform in a non-trivial representation of the rotation group now. Restricting to terms  with one derivative per field, it turns out impossible to build an axionic theory without tree level particle production, or even one reproducing the phase shift (\ref{eis}) at the level of the leading order $2\to 2$ scattering. The only way out is to introduce a cubic axion self-interaction with one less derivative,
\be
\label{SAAA}
S_{AAA}=g_{3,1}\int \epsilon^{\alpha\beta}\Tr A\d_\alpha A\d_\beta A\;.
\ee
By combining this vertex with a cubic vertex coming from (\ref{SA}) (which has two extra derivatives) one may then reproduce the correct leading order $2\to 2$ amplitude. At first sight this is not much of a remedy though. Indeed, 
(\ref{SAAA}) is a marginal $\ell_s$-independent axion self-interaction, so one may worry that it gives rise to $\ell_s$-independent particle production in the scattering processes involving axions, which is not even suppressed at low energies. Even worse,  two-dimensional theories of massless scalar particles with marginal self-interactions of this kind typically suffer from nasty IR divergences. So starting with section~\ref{sec:free} we set $\ell_s=0$ and focus on the leading order axionic self-interactions. 

The chance to proceed is related to the following peculiar property of the coupling (\ref{SAAA}). The three particle amplitude corresponding to (\ref{SAAA}) identically vanishes on-shell, even if the momenta of external particles are allowed to take complex values. Conventionally, this implies that the corresponding coupling can be removed by a local field redefinition. It is straightforward to see that no such field redefinition exists for the coupling (\ref{SAAA}). This opens a route to construct axionic strings at general $D$ (at least in the strict $\ell_s=0$ limit) by supplementing (\ref{SAAA}) with an infinite number of higher order in $A$ marginal vertices in such a way that {\it all} tree level amplitudes vanish and no IR divergences arise. This is the main goal of the present paper. We achieve this goal in section \ref{sec:free}. This is done by generalizing an inductive procedure allowing one to build classically integrable actions based on a clever use of multi-Regge limits as presented recently in \cite{Gabai:2018tmm} (see also \cite{Bercini:2018ysh}; some of the early work can be found in \cite{Arefeva:1974bk,Braden:1991vz,Dorey:1996gd}). Imposing that all tree level amplitudes vanish allows one to uniquely  fix all higher order terms in the axionic action in terms of the $g_{3,1}$ coupling introduced in (\ref{SAAA}). In section~\ref{sec:WZW} we take a closer look at the resulting ``pseudofree" theory
and recognize that in this roundabout way we arrived at a very well-known model---the $O(D-2)$ Wess--Zumino--Witten (WZW) theory \cite{Witten:1983ar} at the scale invariant point. Its rank $k$ (or, equivalently, the value of $g_{3,1}$) remains undetermined in the strict $\ell_s=0$ limit. To fix $k$ as well as  $Q_A$,  one needs to revisit one loop interactions between axions $A^{ij}$ and Goldstones $X^i$ at finite $\ell_s$. 
We postpone this calculation untill a separate publication. In the concluding section \ref{sec:future} we explain why this is more subtle at general $D$ as compared to the $D=4$ case and discuss  future directions.

\section{Axionic Strings at General $D$: Preliminaries}
\label{sec:symmetry}
A straight infinitely long string spontaneously breaks the bulk Poincar\'e group $G=ISO(1,D-1)$ down to 
\[
H=ISO(1,1)\times O(D-2)\;.
\]
 A systematic recipe to build a general low energy effective action describing such a system is provided by the Callan--Coleman--Wess--Zumino (CCWZ) construction \cite{Coleman:1969sm,Callan:1969sn} or more precisely, by its generalization \cite{Isham:1971dv,Volkov:1973vd} to spontaneously broken space-time symmetries (see, e.g., \cite{Delacretaz:2014oxa}, for a recent user friendly introduction). The theory is guaranteed to contain massless Goldstone modes $X^i$ associated with the
spontaneous breaking of space-time translations. In addition to the shift  and $O(D-2)$ rotational symmetries,
\[
X^i\to O^i_j X^j+a^i
\]
they enjoy a symmetry under non-linearly realized off-diagonal boosts/rotations $J_{\alpha i}$, which act as
\be
\delta_{\alpha i} X^j=-\l\delta^{ij}\sigma_\alpha
+X^i\d_\alpha X^j\r\,.
\label{dx}
\ee
Recent reviews of the effective string theory covering the case when $X^i$'s are the only light fields on the worldsheet can be found in \cite{Dubovsky:2012sh,Aharony:2013ipa}.
In particular, the leading order interactions of $X^i$'s are governed by the Nambu--Goto action,
\be
S_{NG}=\ell_s^{-2}\int\sqrt{-\det{\l\eta_{\alpha\beta}+\d_\alpha X^i\d_\beta X^i\r}}\;,
\ee
where 
\[
\eta_{\alpha\beta}=\l\begin{array}{cc}
0 & -1\\
-1 & 0
\end{array}\r\;
\]
is the flat Minkowski metric, and we always work in the light cone coordinates $(\sigma^+,\sigma^-)$. In what follows we also use the convention 
\[
\epsilon^{+-} =-\epsilon^{-+} =1\;.
\]

Additional fields in the CCWZ formalism  are characterized by their quantum numbers w.r.t.  the unbroken subgroup $H$. In axionic strings one introduces a set of worldsheet scalars $A^{ij}=-A^{ji}$ transforming as an antisymmetric $O(D-2)$ tensor. The CCWZ construction provides a systematic way to work out their transformations under the non-linearly realized generators $J_{\alpha i}$. In  Appendix~\ref{CCWZ} we sketch 
this procedure  (a recent detailed discussion of the analogous procedure for effective strings carrying fermionic worldsheet degrees of freedom can be found in \cite{Mohsen:2016lch}).
The resulting transformations of $A^{ij}$ take the following form,
\be
\delta_{\alpha i} A^{kl}\approx -X^i\d_\alpha A^{kl}+
{1\over 2}\d_\alpha X^j\l\delta^{ik} A^{jl}-\delta^{il} A^{jk} \r-{1\over 2}\d_\alpha X^k A^{il}+{1\over 2}\d_\alpha X^l A^{ik}
+\dots
\label{da}
\ee
This transformation law is approximate, because (unlike in (\ref{dx})), we dropped higher order terms in $X$ here, as indicated by  dots.
As a check, note that commutators of these transformations satisfy the target space Lorentz algebra at the leading order in $X$. As an additional check, restricting to $D=4$ and plugging in
\[
A^{ij}=a\epsilon^{ij}\;,
\]
one obtains that only the first term in the r.h.s. of (\ref{da}) survives, giving the correct transformation law for a (pseudo)scalar
\[
\delta_{\alpha i}a=-X^i\d_\alpha a\;.
\]

The leading order term in the $A^{ij}$ action  is their kinetic term
\be
S_{AA}=-\int \frac{1}{4}(\d_\beta A^{kl})^2\,.
\label{SAA}
\ee
A variation of this term under (\ref{da}) takes the following form,
\be
\label{dkin}
\delta S_{AA}={1\over 2}\int\l\d_\beta A^{kl}\d_\alpha A^{kl}\d^\beta X^i-{1\over 2}(\d_\beta A^{kl})^2\d_\alpha X^i+\d_\alpha\d_\beta X^k\l\d_\beta A^{kl} A^{il}-\d_\beta  A^{il} A^{kl}\r
%-\d_\beta  A_{i}^{\;\,l}\d_\alpha\d_\beta X^k A_{kl}+\d_\beta A_{kl}\d_\alpha\d_\beta X^k A_i^{\;\,l} 
 \r\;.
\ee
Note that the first pair of terms in (\ref{dkin}) has a different flavor structure as compared to the remaining two terms. The first two can be cancelled by the variation of the following quartic vertex,
\be
\label{SAAXX}
S_{AAXX}={1\over4}\int \d_\beta A^{kl}\d_\alpha A^{kl}\d^\beta X^i\d^\alpha X^i-{1\over 2}(\d_\beta A^{kl})^2(\d_\alpha X^i)^2\;.
\ee
This vertex has the same form as one would get from expanding the Nambu--Goto action describing $X^i$, $\ell_sA^{kl}$ on equal footing. It is straightforward to check that its variation under (\ref{dx}) indeed cancels the first two terms in (\ref{dkin}). Also this vertex on its own gives the desired $XXAA$ amplitude at this order,
so we need to make sure that no additional contributions arise.
To cancel the remaining two terms in (\ref{dkin}) we need to introduce a quartic vertex of the form
\be
S_{XAAX}={1\over 2}\int\d_\beta A^{kl}A^{il}\l \d_\alpha X^i\d^\alpha\d^\beta X^k-\d_\alpha X^k\d^\alpha\d^\beta X^i\r\;.
\ee
This vertex gives rise to the amplitude of the form
\be
\label{XA4}
{\cal M}_{XAAX}=-{i\ell_s^2\over 2} (p_1p_2)((q_1-q_2)(p_1-p_2)) X_1(p_1)[A_1(q_1),A_2(q_2)]X_2(p_2)\;,
\ee
where $X_1(p_1),X_2(p_2),A_1(q_1),A_2(q_2)$ are the flavor wave functions of colliding particles with incoming momenta $p_1$, $p_2$, $q_1$, $q_2$, and we suppressed all flavor and tensor indices\footnote{Here and in what follows, whenever flavor indices are suppressed, the matrix notation is adopted. For instance, in (\ref{XA4})  \[X_1[A_1,A_2]X_2\equiv X^i_{1}(A_1^{ij}A_2^{jk}-A_2^{ij}A_1^{jk})X^k_2\;.\]}.
This amplitude needs to get canceled for integrable axionic strings.

\begin{figure}[t!]
  \begin{center}
        \includegraphics[width=16cm]{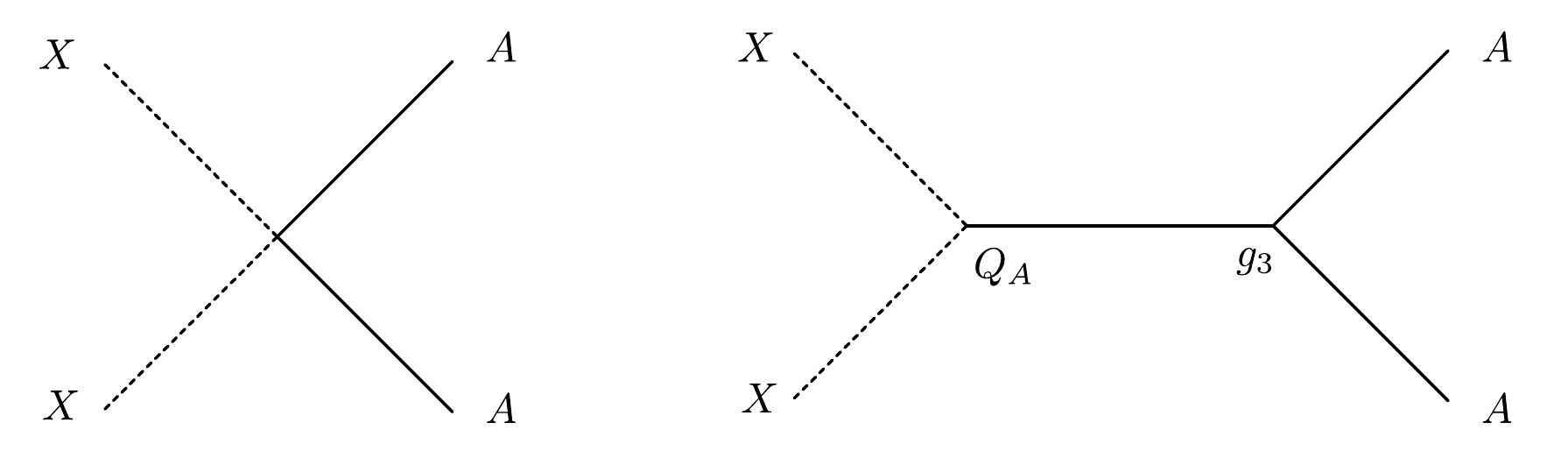} 
           \caption{A cubic axion self-interaction (\ref{SAA}) combined with the $AXX$ vertex (\ref{SAXX}) added together with the contact quartic amplitudes allows to reproduce the correct structure of the $XXAA$ amplitude.}
        \label{fig:4+3}
    \end{center}
\end{figure}

Note that  there are two additional quartic vertices  invariant under Galilean  shifts $X^i\to X^i+\sigma^\alpha$, and hence unconstrained by the non-linearly realized Poincar\'e symmetry,
\be
\label{Sinv}
S_{inv}=\int C_1\d_\alpha\d_\beta X^iA^{ik}A^{kj}\d^\alpha\d^\beta X^j+C_2(\d_\alpha\d_\beta X^i)^2(A^{kl})^2\;.
\ee
The second vertex in (\ref{Sinv}) has the same flavor structure as (\ref{SAAXX}), so it should vanish, $C_2=0$, for integrable axionic strings.
The first vertex gives rise to the amplitude of the form
\be
{\cal M}_{inv}=2C_1i\ell_s^2 (p_1p_2)^2X_1(p_1)\{A_1(q_1), A_1(q_2)\}X_2(p_2)\;,
\ee
which is different from the one in (\ref{XA4}). 
This implies that the only chance to cancel the amplitude (\ref{XA4}) is to set $C_1=C_2=0$ and to introduce a lower order in derivative cubic self-interaction of axions (\ref{SAAA}).
 Combining this vertex with the axionic $AXX$ interaction
 \be
 \label{SAXX}
 S_{AXX}=Q_A\int \epsilon^{\alpha\beta}\d_\alpha\d_\gamma X^i A^{ij}\d_\beta\d^\gamma X^j\;,
 \ee
 coming from expanding (\ref{SA}) to the leading order in $X$,
 one obtains an additional ${\cal O}(\ell_s^2)$ contribution to the $XAAX$ amplitude. As shown in Fig.~\ref{fig:4+3}, this contribution may be used to cancel (\ref{XA4}).
 Indeed, the corresponding amplitude is
 \be
 \label{AAAAXX}
 {\cal M}_{AAA,AXX}=-6ig_{3,1}Q_A\ell_s^2((p_1 q_1)^2-(p_1 q_2)^2)X_1(p_1)[A_1(q_1),A_2(q_2)]X_2(p_2)\;.
 \ee
 This amplitude has the same flavor structure as (\ref{XA4}) and it is straightforward to check that for all on-shell configurations of the massless momenta it is indeed proportional to (\ref{XA4}). Hence, we can choose the value of $g_{3,1}$ in such a way that (\ref{AAAAXX}) cancels against (\ref{XA4}), namely
 \be
 \label{naiveg31}
 g_{3,1}={1\over 6 Q_A}\;.%={1\over 3}\sqrt{12\pi\over 26-D-(D-2)(D-3)/2}\
 \ee
 This completes the construction of the leading order axionic string action that reproduces the tree level $XXAA$ amplitude in agreement with (\ref{eis}).
 
\section{Pseudofree Axions}
\label{sec:free}
We see that  $D>4$ integrable axionic strings, if they exist,  necessarily have a marginal cubic interaction of  the form (\ref{SAAA}). This interaction survives even in the $\ell_s\to 0$ limit, which is somewhat surprising given that the worldsheet $S$-matrix (\ref{eis})
becomes trivial in this limit. In the rest of the paper we will study  the resulting axion ``self-interactions" at $\ell_s=0$. 
We will see that the above contradiction gets resolved in a rather interesting way.
The cubic vertex (\ref{SAAA}) exhibits the following unconventional property. It identically vanishes on-shell even if particle momenta are analytically continued in the complex domain. Normally, interaction vertices with such a property can be removed from the action by a local field redefinition. It is straightforward to check that this is impossible in the present case. Hence, for integrable axionic strings to exist we need to show that higher order two-derivative axion self-interactions can be introduced in such a way that the axion $S$-matrix stays trivial at $\ell_s=0$. If it exists, the resulting $\ell_s=0$ theory is quite peculiar---it has a trivial $S$-matrix, however, its action cannnot be brought into a free form by a local field redefinition. It is natural to refer to a theory
with this property as a pseudofree one. 

Let us start with a straightforward inductive argument demonstrating that the pseudofree theory can indeed be constructed. We will limit our analysis to tree level. Note first that the structure of  axion self-interactions is not restricted by the non-linearly realized symmetry, given that the transformation rule (\ref{da}) necessarily involves Goldstone fields (and also increases the number of derivatives). Then to construct the leading order axionic  Lagrangian one needs to calculate axionic scattering amplitude with larger and larger number of external legs. The cubic amplitude is determined by (\ref{SAAA}) and vanishes, providing the base for the inductive argument. To prove the inductive step, assume that we managed to construct the action including axionic vertices with up to $n$ legs, such that all amplitudes involving $n$ or smaller number of axions vanish.
Then the amplitudes with $(n+1)$ external legs following from this action cannot have any factorization poles. Hence it can be cancelled as well with 
an appropriate choice of a local axionic vertex with $(n+1)$ legs, which completes the proof. 

After a pseudofree theory is built at the tree level, there should be no obstruction to extend the construction at an arbitrary loop order.
Indeed, the singularities of higher loop amplitudes are fixed by lower order amplitudes. Hence if all lower loop amplitudes are trivial one should be able to extend the theory  at the next order in the loop expansion.

However, one may worry that this reasoning is too fast and may be spoiled by IR divergences. Indeed, the axion self-interaction (\ref{SAAA}) contains an axion field without any derivative acting on it, which usually implies the presence of IR divergencies in two dimensions. So, to eliminate these concerns, in the rest of this section we elaborate on this argument and will explicitly follow through the tree level inductive procedure. This will allow us to derive  a set of recursion relations on the axion couplings, which completely fix 
the form of the action. The corresponding analysis is somewhat technical though and an impatient reader, who trusts our skills in manipulating tree level Feynman diagrams, may skip directly to the final result (\ref{recursions}).  

\subsection{General Structure of the Lagrangian and Feynman Rules}
To set the stage let us describe a convenient way to organize Feynman rules in the axionic theory. 
Given that our seed cubic vertex (\ref{SAAA}) is a single trace operator, one expects also the full pseudofree Lagrangian to be a sum of single trace operators to all orders in $A$. Hence, we are led to search for the axion action in the form
\begin{gather}
S_A=S_{AA}+\int \sum_{J=2}^\infty\l\sum_{m=1}^{J-1}g_{2J,m}\eta^{\alpha\beta}\Tr\,{\partial_\alpha A A^{m-1}\partial_\beta A A^{2J-m-1}}%\nonumber\\
+ \frac{g_{2J,J}}{2}\eta^{\alpha\beta}\Tr\,{\partial_\alpha A A^{J-1}\partial_\beta A A^{J-1}}\r\nonumber\\
+  \int \sum_{J=1}^\infty\sum_{m=1}^{J}g_{2J+1,m}\epsilon^{\alpha\beta}\Tr\,{\partial_\alpha A A^{m-1}\partial_\beta A A^{2J-m-1}}\;,
\label{gaction}
\end{gather}
where the free action $S_{AA}$ is given by (\ref{SAA}).
Taking into account the cyclic property of the trace, (\ref{gaction}) includes all possible 
parity invariant marginal axion self-interactions. An extra factor of $1/2$ in the definition of $g_{2J,J}$ couplings is included here for the later convenience.

Normally, for matrix theories like (\ref{gaction}) 't Hooft double-line notations \cite{'tHooft:1973jz}
provide a convenient way to keep track of the flavor factors. However, axions belong to the orthogonal algebra $O(D-2)$ rather than to a unitary one, so that double-line notations are not directly applicable. Nevertheless, the counting of flavor factors  for the orthogonal  
groups is also governed by topology, although one needs to allow for non-orientable surfaces as well \cite{Cicuta:1982fu}. In particular, at the leading order in $1/D$ expansion there is no difference between orthogonal and unitary groups \cite{Lovelace:1982hz,2008JPhA...41L2001Z}. 
 Even though  we are not performing the $1/D$ expansion here, our analysis is restricted to tree level, where no subleading contributions in $1/D$  arise.
 In particular, just like in the unitary case, any tree level amplitude can be written in the form
 \begin{equation}
 \label{Mc}
\mathcal{M}\left(\{p^1,T^1\}\cdots\{p^n,T^n\}\right) = \sum_{\sigma\in P_n/Z_n}\Tr\,{T^{\sigma(1)}\cdots T^{\sigma(n)}}\mathcal{M}_c(p^{\sigma(1)},\cdots p^{\sigma(n)})\;,
\end{equation}
where $\{p^i,T^i\}$ are the momenta and $O(D-2)$ polarizations of  colliding axions. The sum in (\ref{Mc}) is performed over all permutations which are not related by cyclic reorderings. Color-ordered amplitudes $\mathcal{M}_c(p^{1},\cdots, p^{n})$ can be calculated using color-ordered Feynman rules very similar to those in the unitary case (c.f. \cite{Dixon:1996wi}). Namely, one needs to 

{\bf 1.} Draw all tree graphs with $n$ external legs and without self-intersections, where the cyclic ordering of external momenta matches the one in $\mathcal{M}_c$.

\begin{figure}[t!]
  \begin{center}
        \includegraphics[width=10cm]{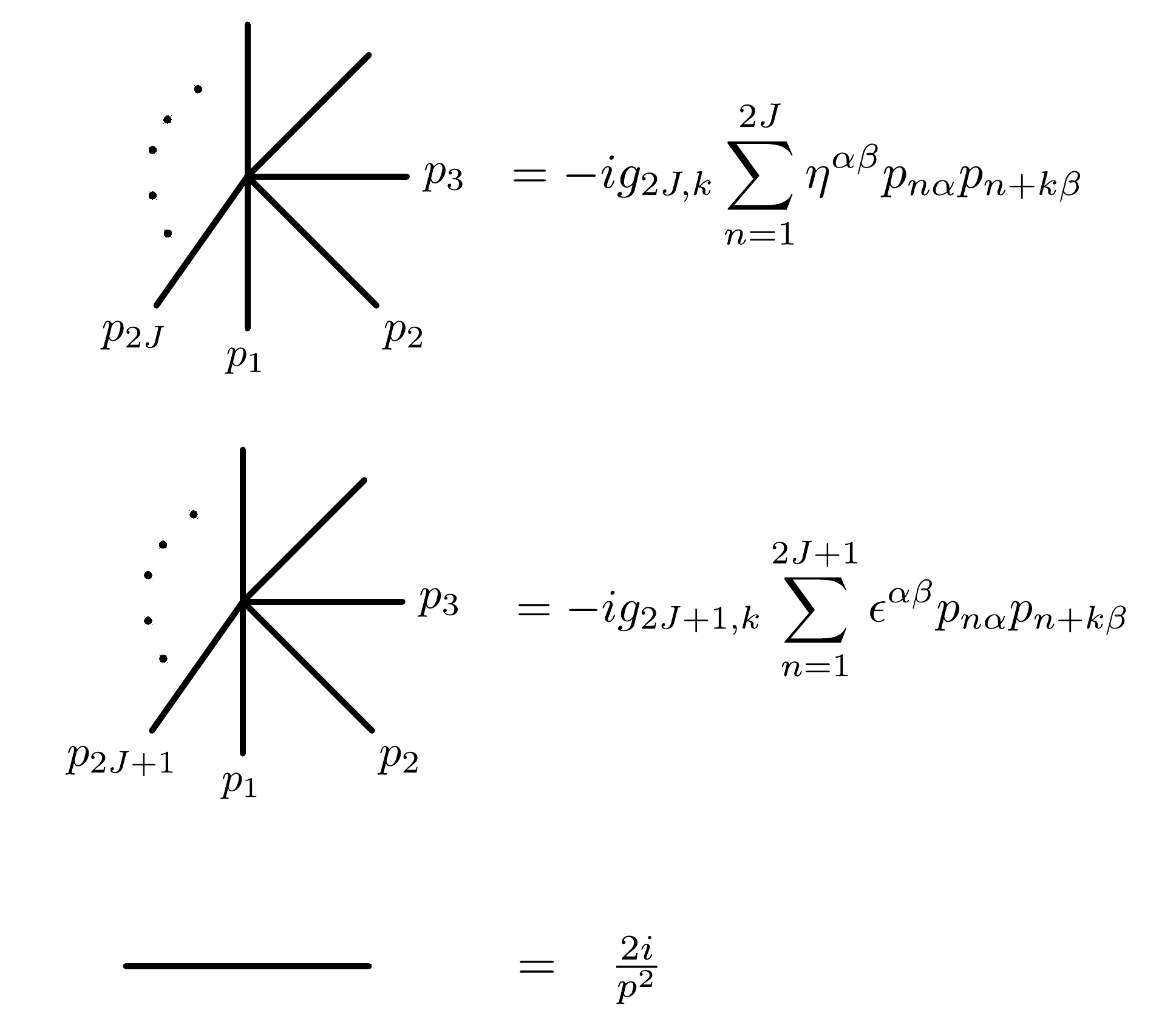} 
           \caption{Feynman rules for color-ordered amplitudes.}
        \label{fig:Feynman}
    \end{center}
\end{figure}

{\bf 2.} Evaluate each graph using the vertices and propagators of Fig.~\ref{fig:Feynman}.

Note that a two-point function of the axion fields following from (\ref{SAA}) is equal to
\begin{equation}
\langle A^{ij}(p)A^{kl}(-p) \rangle = -\frac{i}{p^2}\left(\delta_{ik}\delta_{jl} - \delta_{il}\delta_{jk}\right)
\end{equation}
which translates into 
 an extra factor of 2 in the color-ordered propagator in Fig.~\ref{fig:Feynman}. 
 
 \subsection{Absence of IR divergences}
 \label{subsec:IR}
Before  going into details of the inductive procedure, which generalizes the one presented in  \cite{Gabai:2018tmm} and makes it possible to fix all $g_{n,k}$ couplings in terms of $g_{3,1}$, let us comment on possible IR divergences (a detailed discussion of these issues in a very close context just appeared in \cite{Hoare:2018jim}). 
Generically, theories of massless particles in two dimensions are plagued with  IR divergences. Physically, these arise because a bunch of massless left- (or right-) movers emitted from an interaction region never get spatially
separated in a linear kinematics, so that no asymptotic states can be defined. A notable exception occurs when a massless theory at low energies flows into a free CFT perturbed by a set of irrelevant operators (cf. \cite{Zamolodchikov:1991vx,Dubovsky:2012wk}). 

However,  all interactions in (\ref{gaction}) are marginal, so a priori one expects to find IR issues in the corresponding on-shell amplitudes.
Indeed, these were encountered in the analysis of   \cite{Gabai:2018tmm} (see also \cite{Hoare:2018jim} for a dedicated discussion). There the goal was to construct an integrable model for an $U(N)$ analogue of (\ref{gaction}) with all odd couplings set to zero, $g_{2J+1,m}=0$.
It turned out possible to rediscover  an integrable non-linear $U(N)$ sigma model by requiring  that IR safe multiparticle amplitudes, such as ${\mathcal M}_c(+-\dots+-)$, vanish\footnote{Here and in what follows, $+$'s and $-$'s show whether a corresponding momentum is left- or right-moving. Also, we treat all momenta as incoming.}. However, some other amplitudes in \cite{Gabai:2018tmm}, such as ${\mathcal M}_c(+\dots+-\dots-)$ remain non-zero even in an integrable theory. This  indicates the presence of IR ambiguities.

As we will see now, the situation in a pseudofree case is different. Namely, it is possible to find a set of coupling constants $g_{n,m}$ in (\ref{gaction}) such that {\it all} on-shell scattering  amplitudes vanish, including ones which do not correspond to IR safe kinematics. Indeed, as the physical argument above indicates the trouble is caused by scattering of left- (or right-) movers off each other. For instance, on-shell  diagrams involving only left-movers ${\mathcal M}_c(+\dots+)$, are not well-defined at face value.
On one side they look singular, because all  internal propagators are on-shell, on the other hand all interaction vertices in these diagrams vanish as well.

However, these diagram on their own don't cause much trouble. It is natural to try to define a massless theory as a limit of a massive one.
Upon taking the $m^2\to 0$ limit in such a way that all external particles become left-movers any amputated tree level  diagram scales
as
\[
{\mathcal M}_c(+\dots+)\propto m^{2(V-P)}=m^2\;,
\]
where $V$ is the number of vertices, and $P$ is the number of internal propagators. So it is natural to set all amplitudes of this kind to zero in the massless limit.

\begin{figure}[t!]
  \begin{center}
        \includegraphics[width=10cm]{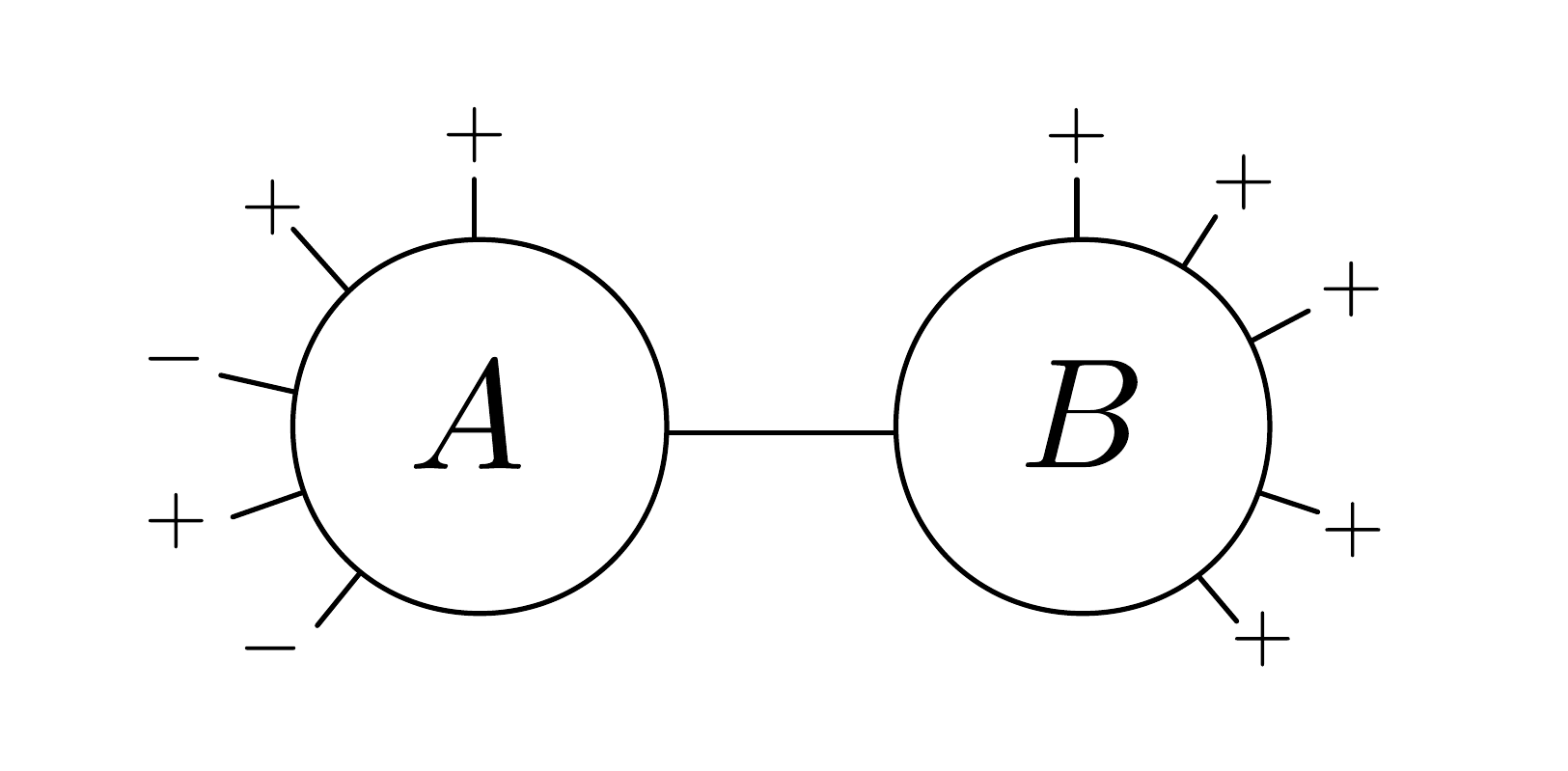} 
           \caption{IR ambiguities are associated with Feynman diagrams containing subdiagrams where all massless particles move in the same direction.}
        \label{fig:problematic}
    \end{center}
\end{figure}

Instead, the real problem arises when an amplitude like that arises as a subdiagram in a process with a larger number of particles and gets attached to some non-trivial amplitude, see Fig.~\ref{fig:problematic}. In this case upon taking the massless limit one obtains an additional singular propagator---the one connecting the purely left-moving subdiagram $B$ with the rest. As a result, contributions like this stay finite at $m^2= 0$, which looks unphysical. For instance, in general this limit looks ambiguous, because the result depends on the mass ratios   of different particles as their masses are being taken to zero. Indeed,  the non-linear sigma model avoids these ambiguities 
by leading to factorized scattering of massive particles at the end of the day.

However, the situation is better in a pseudofree case. Indeed, we will be constructing this theory inductively in the number of colliding particles. Then for the diagram shown
in Fig.~\ref{fig:problematic} the $A$ subamplitude also goes on-shell and hence vanishes in the massless limit. As a result the diagram as a whole vanishes as well. Hence, in this case it is consistent to set to zero all diagrams of this kind and no IR ambiguities arise.

\subsection{Vanishing of $\mathcal{M}_c(+\dots+-\dots-)$} 
\label{subsec:++}
The discussion in section~\ref{subsec:IR} indicates that IR divergencies do not spoil the argument presented in the beginning of the section, and that a pseudofree theory can indeed be constructed.
To construct the corresponding action we follow the strategy of \cite{Gabai:2018tmm}. Namely, we fix all the coefficients $g_{n,m}$ by requiring that a sufficiently large
 subclass of amplitudes vanishes. The evaluation of the corresponding amplitudes  is made tractable by considering convenient kinematical limits. The procedure is inductive in the number of external legs. 
It is immediate to see that the presence of an irreducible cubic vertex in our case does not allow to use the same kinematics as in  \cite{Gabai:2018tmm}.
 Instead, to start with we consider amplitudes with $m$ left-movers and $n$ right-movers ordered according to 
 \[
 \mathcal{M}_c(+\dots+-\dots-)\equiv\mathcal{M}_c(m,n)\;.
 \] 
 These amplitudes do not factorize in the integrable theory constructed in  \cite{Gabai:2018tmm} as a consequence of IR ambiguitites, but as discussed in section~\ref{subsec:IR}, they still have to vanish in a pseudofree theory. We will consider the multi-Regge limit defined by the following choice of momenta
\begin{gather}
p_{j}^+ = E x^j\,,\;\;\;j=1,\dots, m-1\\
p_{j}^- = E y^{j-m}\,,\;\;\;j=m+1\dots m+n-1\;.
\end{gather}
Here $E$ is an arbitrary energy scale, which we set to one in what follows. 
\begin{figure}[t!]
	\centering
	\includegraphics[width=0.5\linewidth]{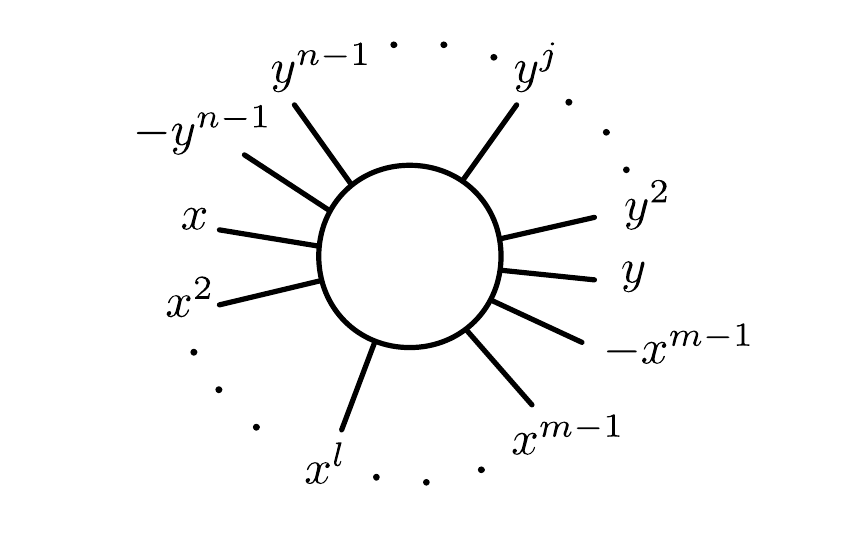}
	\caption{Momentum configuration in the multi-Regge kinematics we consider for $\mathcal{M}_c(m,n)$.}
	\label{fig:highenergypppmm}
\end{figure}
We work in the limit
\be
x,y\gg1
\ee
and  impose that amplitudes vanish at the leading order in $x$ and $y$. Momentum conservation determines the remaining momenta $p^+_{m}$ and $p^-_{n+m}$ to be given by
\begin{align}
p^+_{m} \approx - x^{m-1}+\dots\\\nonumber
p^-_{n+m} \approx -y^{n-1}+\dots\nonumber
\end{align}
where dots stand for subleading terms in $x$ and $y$. In the later formulas and figures the dots will be implied, but not written explicitly.
This configuration of external momenta is shown in figure \ref{fig:highenergypppmm}.

\begin{figure}[t!]
	\centering
	\includegraphics[width=0.99\linewidth]{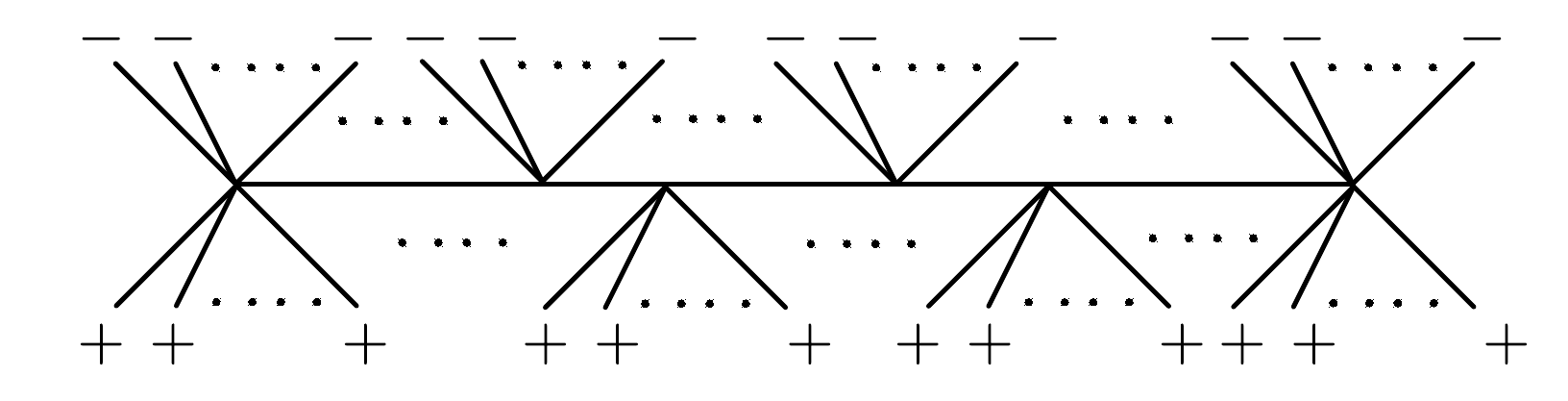}
	\caption{$\mathcal{M}_c(m,n)$ amplitudes receive contributions only from the chain diagrams of the type shown here.}
	\label{fig:generalchain}
\end{figure}

A nice property of  $\mathcal{M}_c(m,n)$ amplitudes is that only chain diagrams shown in Fig.~\ref{fig:generalchain} contribute in this kinematics. Indeed, any non-chain diagram corresponds to a tree with at least three branches. Then at least one of these branches contains either all left-movers or all right-movers, so that the corresponding diagram vanishes according to the argument in section~\ref{subsec:IR} (assuming, for instance, the mass regularization as done there).

\begin{figure}[t!]
	\centering
	\includegraphics[width=0.8\linewidth]{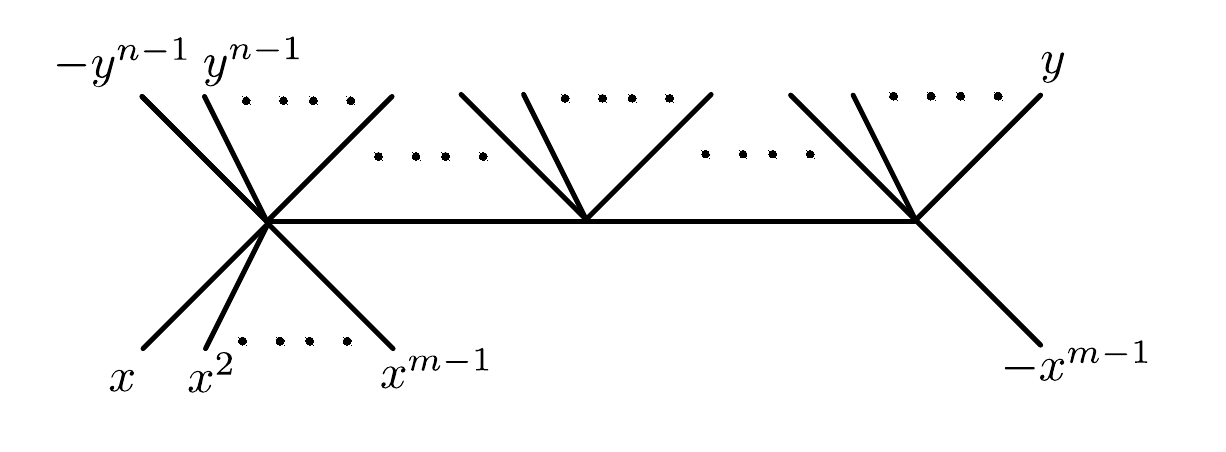}
	\caption{Chain diagrams which contribute into $\mathcal{M}_c(m,n)$ at the leading order at large $x$.}
	\label{fig:maxx}
\end{figure}

\begin{figure}[t!]
	\centering
	\includegraphics[width=0.99\linewidth]{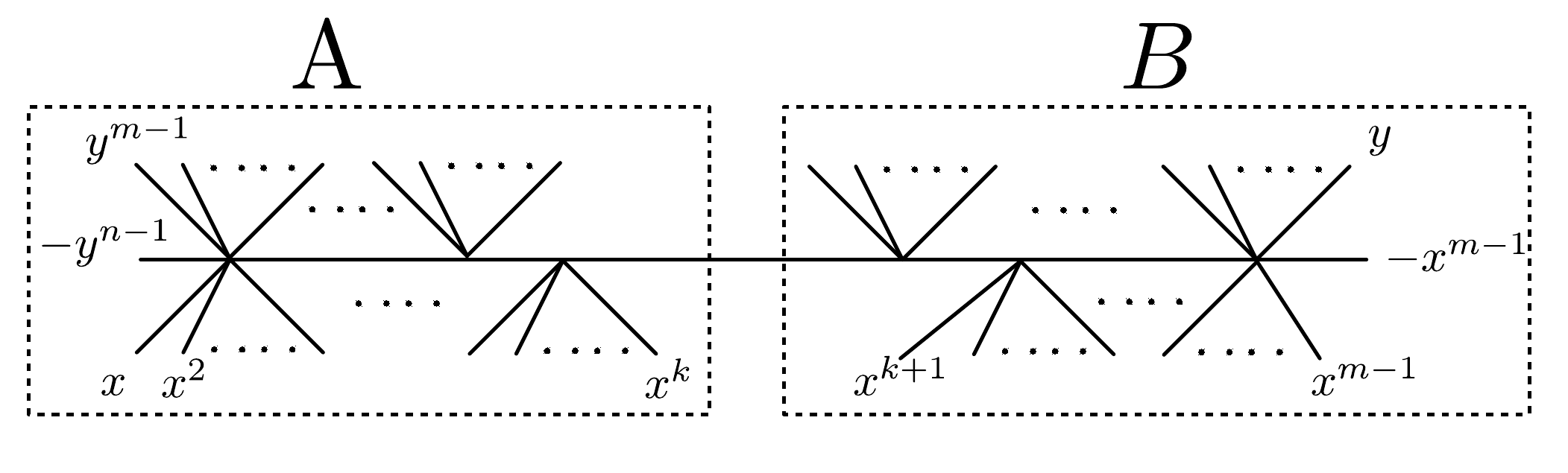}
	\caption{Internal lines in a generic chain diagrams are approximately on-shell in the large $x$ limit. As a result  generic chains  give a subdominant contribution in the multi-Regge kinematics.}
	\label{fig:nonmaxx}
\end{figure}

Suppose now that we managed to set to zero all $\mathcal{M}_c(m_1,n_1)$ amplitudes with less than $(m+n)$ external legs by an appropriate choice of $g_{k,l}$ with $k<m+n$. Then it is straightforward to see that among non-trivial chains with $(m+n)$ legs only the ones where all $x,\dots x^{m-1}$ momenta enter into the left-most vertex, and $-x^{m-1}+\dots$ momentum enters into the right-most vertex contribute at the leading ${\cal O}(x^{m-1})$ order in $x$, see Fig~\ref{fig:maxx}. 
Indeed, for all other chain diagrams there is an internal line such that there is at least two $x$ momenta on the right of it, see Fig~\ref{fig:nonmaxx}. In general this line is not on-shell, but its $x$ momentum is necessarily smaller than all $x$ momenta on the right. Hence, at the leading order in $x$ the subdiagram $B$ on the right is on-shell (with the internal line carrying $y$ momentum). By summing all diagrams of this kind with the same subdiagram $A$ on the left of the internal line we get zero at the leading order in $x$, as a consequence of the inductive assumption.  

\begin{figure}[t!]
	\centering
	\includegraphics[width=1\linewidth]{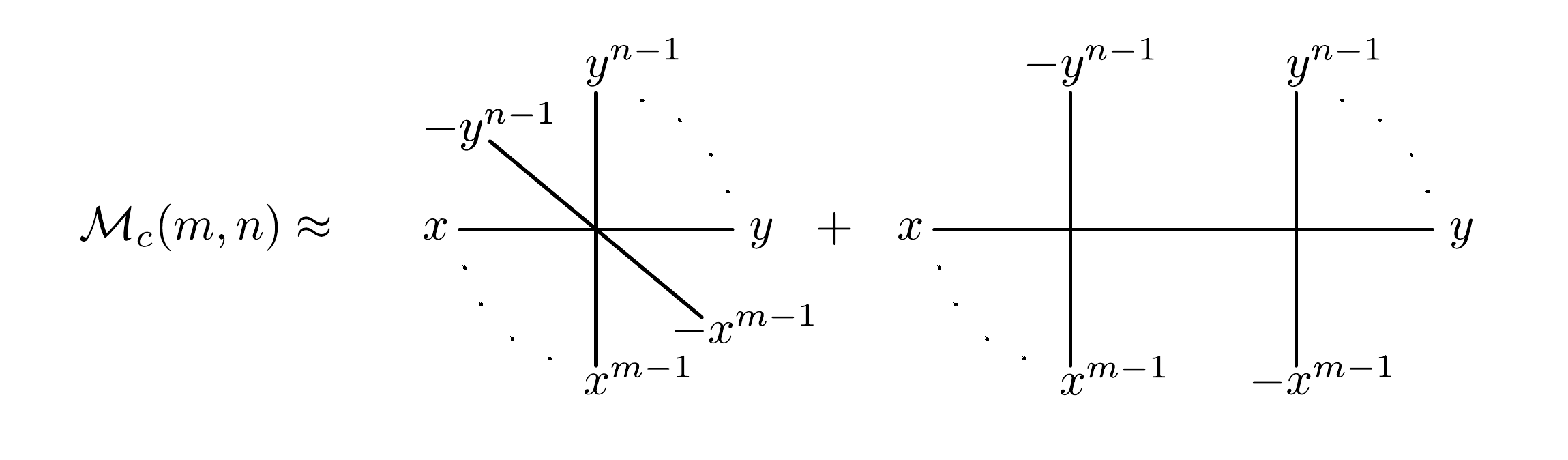}
	\caption{The only two diagrams which contribute into $\mathcal{M}_c(m,n)$ at the leading order in the multi-Regge kinematics.}
	\label{fig:ppmmtreediags}
\end{figure}

The same arguments applies to $y$ momenta. As a result we conclude that at the leading order both in $x$ and in $y$ only two tree level diagrams survive---a chain of length one, and a ``trivial" chain (the contact vertex), 
see  Fig.~\ref{fig:ppmmtreediags}.
It is straightforward to calculate these two diagrams using the Feynman rules shown in Fig.~\ref{fig:Feynman}.
At intermediate steps one needs to treat separately the cases of even and odd $(m+n)$, but the final result can be written in a universal  compact  form. Namely,  at the  leading order at large $x,y$,
the contact contribution turns into 
\be
\label{contodd}
{\cal M}_{contact}(m,n)=i(-1)^{m+n+1}(2g_{m+n,m}-g_{m+n,m-1}-g_{m+n,m+1})x^{m-1}y^{n-1}\;.
\ee 
Note that the  coupling constants $g_{m+n,m}$ in (\ref{gaction}) are only defined for $m\leq n$. In (\ref{contodd}) we defined them also at $m>n$ via 
\be
\label{swap}
g_{m+n,m}\equiv (-1)^{m+n}g_{m+n,n}\;,\;\;\; \mbox{\rm for}\;\;\; m>n\;, 
\ee
where a minus sign for odd $(m+n)$ is due to the   presence of the $\epsilon^{\alpha\beta}$ tensor in a vertex with an odd number of external legs.

Similarly, for the chain contribution at leading order in  $x,y$ one gets
\be
\label{chainodd}
{\cal M}_{chain}(m,n)=i(-1)^{n+1}(2g_{m+1,1}-g_{m+1,2})(2g_{n+1,1}-g_{n+1,2})x^{m-1}y^{n-1}\;.
\ee

By requiring
\be
\label{M+M}
{\cal M}_{contact}(m,n)+{\cal M}_{chain}(m,n)=0\;,
\ee
we obtain a number of recursion relations for the coupling constants $g_{m+n,m}$. However, it is straightforward to see that we need additional relations to fix all  $g_{m+n,m}$'s in terms of $g_{3,1}$.
Indeed, independent couplings are $g_{m+n,m}$ with $1\leq m\leq n$. On the other hand, as a consequence of parity invariance, relations  ${\cal M}(m,n)=0$ with $m>n$  follow  from ${\cal M}(n,m)=0$.
In addition, there is no relation for $m=1$ because those amplitudes vanish trivially. Hence we are missing one relation for each $(m+n)$.

In fact, we are doing slightly better than that.
Indeed, by making use of field redefinitions of the form
\be
\label{redef}
A\to A+\sum_{J=1}^\infty \alpha_J A^{2J+1}\;,
\ee
we may impose one additional constraint on $g_{n+m,m }$ at any even $(n+m)$. This still  leaves us with one undetermined coupling at any odd $n+m>3$. Therefore we do need to impose an additional set of relations.
\subsection{Vanishing of ${\cal M}_c(+-+-\dots-)$}
\begin{figure}
	\centering
	\includegraphics[width=1\linewidth]{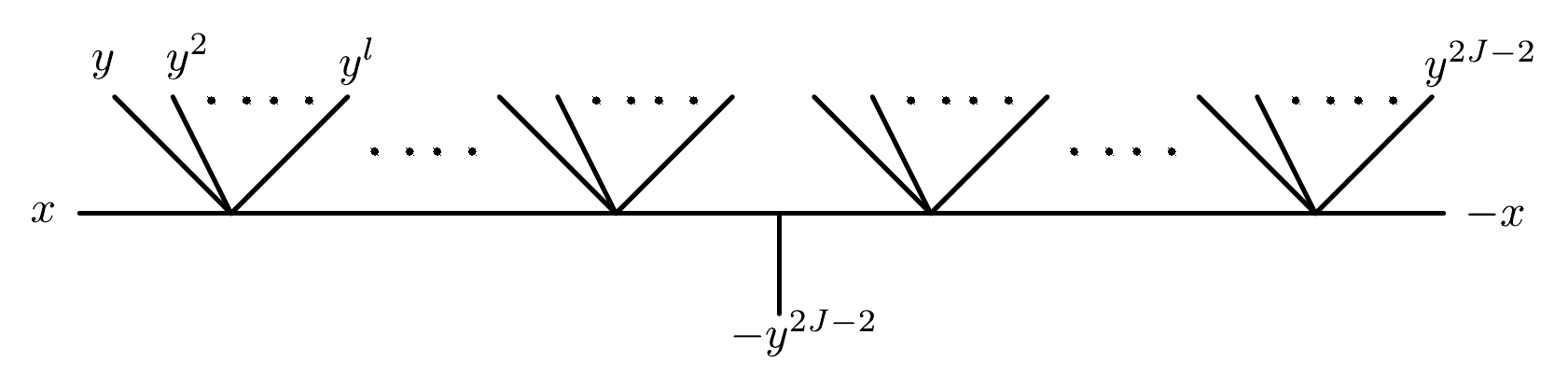}
	\caption{ Chain diagrams contributing into $\mathcal{M}_c(J)$.}
	\label{fig:genericchain}
\end{figure}

To obtain an additional set of recursive relations let us consider  amplitudes with $(2J+1)$ external legs of  the  form
\be
\label{M+-+}
{\cal M}_c(J)\equiv{\cal M}_c(+-+-\dots -)={\cal M}_c(x,-y^{2J-2}+\dots,-x,y^{2J-2},\dots,y)
\ee
at $y\gg 1$.
A nice property of these amplitudes is that, similarly to ${\cal M}_c(m,n)$ considered previously, 
${\cal M}_c(J)$'s are also given by a sum of chains, see Fig.~\ref{fig:genericchain}.
Although this greatly reduces the number of tree-level diagrams involved in the calculation,  a significant number of diagrams still remain. A calculation in the high energy limit greatly simplifies if in addition a field redefinition can be found that sets to zero the leading in $y$ piece of the sum of all chain diagrams of the form (see Fig.~\ref{fig:offshellgenericchain})
\begin{equation}
\mathcal{C}_c(n)\equiv \mathcal{C}_c(\alpha,+,-\dots- ) = \mathcal{C}_c((x,-y^{n}+\dots),-x,y^{n},\dots,y)\;.
\label{offshellchain}
\end{equation}
with $n>1$.
 Here $\alpha$ stands for an off-shell momentum.  The name $\mathcal{C}_c(n)$ emphasizes  that this object is a sum of  chain diagrams only, rather than a full off-shell amplitude. Since this is an off-shell object, in principle  it may be set to zero through a field redefinition. However, we require that this sum of chains vanish (at the leading order in $y$) for both even and odd number of external legs, and naively we don't have enough parameters $\alpha_J$'s in the field redefinition (\ref{redef}) to ensure this.
Luckily, it turns out that the remaining chains vanish automatically.
Namely, let us prove that the 
 following  iterative procedure can be consistently implemented:
\begin{figure}[t!]
	\centering
	\includegraphics[width=\linewidth]{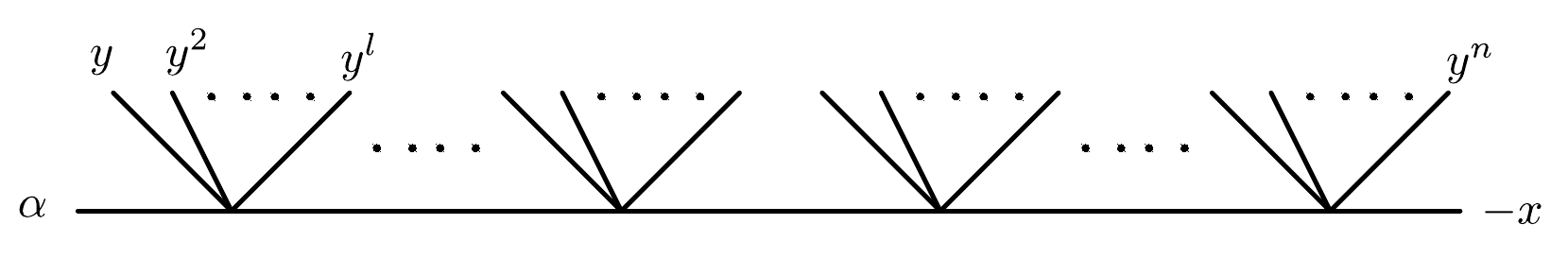}
	\caption{Off-shell chains contributing into $\mathcal{C}_c(n)$.}
	\label{fig:offshellgenericchain}
\end{figure}

\textbf{1.} For a given $J$ assume that couplings $g_{k,m}$ with $k\leq 2J-1$ are chosen in such a way that all on-shell  amplitudes and all $n>1$ chains $\mathcal{C}_c(n)$ 
with $(2J-1)$ or smaller number of external legs vanish (for chains only at the leading order in $y$).

\textbf{2.} Impose that  the $\mathcal{C}_c(2J-2)$  chain vanishes at the leading order in $y$. This gives a new condition for the $g_{2J,k}$ couplings that can be satisfied by making use of a field redefinition of the form
\[
A\to A+\alpha_{J-1}A^{2J-1}\;,
\]
with an appropriately chosen $\alpha_{J-1}$.
% In combination with the $(J-1)$ equations in \eqref{Jm1evenequations} this completely fixes the $g_{2J},k$ coefficients in terms of lower level coefficients.\\
%\\

\textbf{3.} Impose that $\mathcal{M}_c(J)$  vanishes at the leading order in $y$. This gives an additional equation which $g_{2J+1,k}$ couplings must satisfy. 

\textbf{4.} Check that the new conditions obtained in steps \textbf{2} and  \textbf{3} combined with (\ref{M+M})
 automatically imply that $\mathcal{C}_c(2J-1)=0$ at the leading order in  $y$. This ensures that the iterative procedure is consistent and can be carried over to a  larger number of legs.

\begin{figure}[t!]
	\centering
	\includegraphics[width=\linewidth]{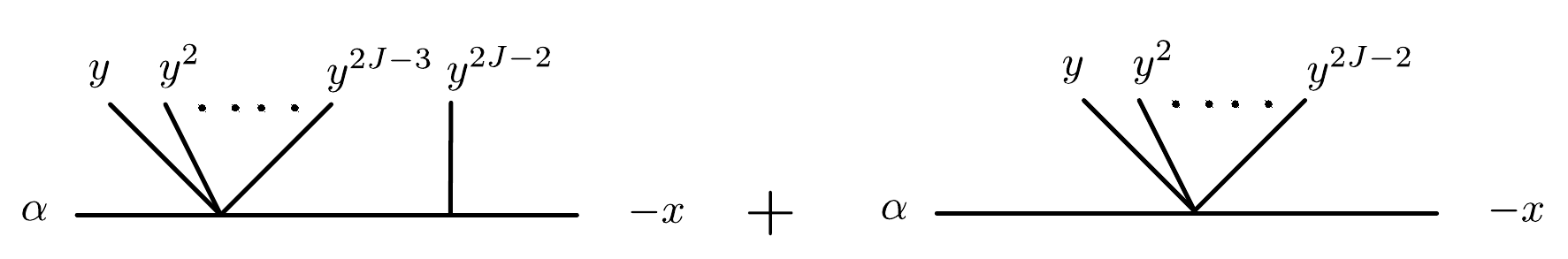}
	\caption{The only two chains contributing to $\mathcal{C}_c(2J-2)$ under the assumption that $\mathcal{C}_c(n)=0$ at $1<n<2J-2$.}
	\label{fig:evenchains}
\end{figure}

To carry out  step \textbf{2}, note that as a consequence of  \textbf{1}
it is only  the two chains shown in Fig.~\ref{fig:evenchains} that contribute to $\mathcal{C}_c(2J-2)$. 
Moreover, it is straightforward to check that at the leading order in $y$ only the right diagram in Fig.~\ref{fig:evenchains} contributes  because of the antisymmetry of the odd vertex, resulting in the condition
\begin{equation}
g_{2J,2} = 0\;.
\label{fieldredefg2J}
\end{equation}

\begin{figure}[t!]
	\centering
	\includegraphics[width=1\linewidth]{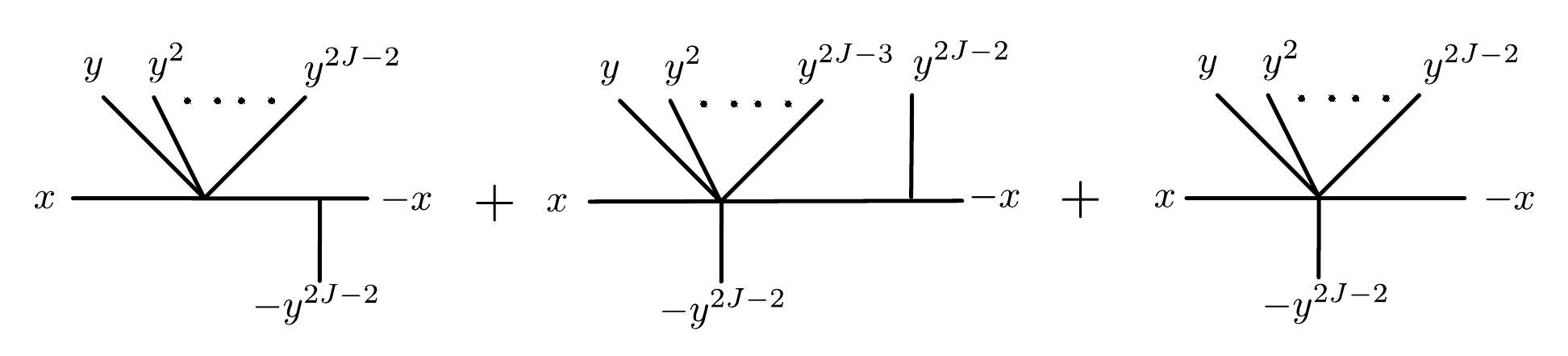}
	\caption{Diagrams conrtibuting to $\mathcal{M}_c(J)$.}
	\label{fig:pmpmamplitude}
\end{figure}

At the step \textbf{3} we need to set $\mathcal{M}_c(J)$ to zero at the leading order in $y$. As a consequence of  \textbf{1}, one is left with three diagrams shown in figure \ref{fig:pmpmamplitude}. 
To the leading order in $y$ these yield
\be
\mathcal{M}_c(J) = ixy^{2J-2}(-6g_3g_{2J,3} + 3g_{2J+1,1} -g_{2J+1,3})=0\;.
\label{step3amp}
\ee

\begin{figure}[t!]
	\centering
	\includegraphics[width=0.75\linewidth]{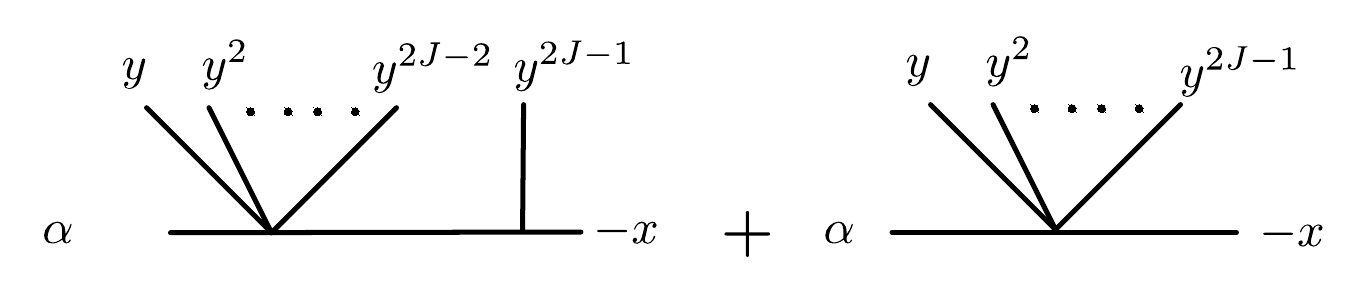}
	\caption{Diagrams conrtibuting to $\mathcal{C}_c(2J-1)$.}
	\label{fig:oddchain}
\end{figure}

Finally, we need to check that $\mathcal{C}_c(2J-1)=0$ at the leading order in  $y$ (step  \textbf{4}). The diagrams that contribute are shown in Fig.~\ref{fig:oddchain} and  the result is 
\be
\mathcal{C}_c(2J-1)= ixy^{2J-1}(-6g_{2J,1}g_3 + 2g_{2J+1,1} -g_{2J+1,2})=0\;.
 \label{step4amp}
 \ee
To see that (\ref{step4amp}) indeed vanishes, note that combining (\ref{fieldredefg2J}) with (\ref{M+M}) applied to ${\cal M}_{c}(2,n)$ gives
\begin{equation}
\label{combo}
2g_{2J+1,2} - g_{2J+1,1} -g_{2J+1,3} +6g_{3}g_{2J,1} = 0\;,
\end{equation}
which, taking into account  \eqref{step3amp}, implies \eqref{step4amp}.
As a result, (\ref{fieldredefg2J}) and (\ref{step4amp}) provide us with one additional constraint on the coupling constants $g_{n,m}$ for any number of external legs $n>3$, which is exactly what is needed to fix all these couplings in terms of $g_{3,1}$.
\section{Reuniting with WZW}
\label{sec:WZW}
To summarize, (\ref{M+M}),  (\ref{fieldredefg2J}) and \eqref{step4amp} give us the following set of recursion relations on the coupling constants $g_{n,m}$
\begin{gather}
\label{recursions}
2g_{n,m}-g_{n,m-1}-g_{n,m+1}+(-1)^m(2g_{m+1,1}-g_{m+1,2})(2g_{n-m+1,1}-g_{n-m+1,2})=0\\
g_{2J,2}=0
\label{rec2}\\
6g_{3,1}g_{2J,3}-3g_{2J+1,1}+g_{2J+1,3}=0\;.
\label{rec3}
\end{gather}
Here (\ref{recursions}) hold for all $n\geq 4$ and $2\leq m\leq\lfloor n/2\rfloor$ with the convention (\ref{swap}) applied when necessary. Relations (\ref{rec2}) and (\ref{rec3}) hold for $J\geq 2$.

Given that we have enough relations to fix all couplings in terms of $g_{3,1}$ it suffices to guess the solution to (\ref{recursions})-(\ref{rec3}) and check it afterwards. This is exactly what we did, using numerical {\it Mathematica} results as well as a solution to a similar problem presented in \cite{Gabai:2018tmm} as a guidance. This leads to the following expression for the coupling constants,
\begin{gather}
g_{2J,2k}=0\;,\;\;1\leq 2k\leq J\label{even1}\\
g_{2J,2k+1}={1\over 2}(3 g_{3,1})^{2(J-1)}\;,\;\;1\leq 2k+1\leq J\label{even2}\\
g_{2J+1,2k}=-{k\over (2J+1)}(3g_{3,1})^{2J-1}\;,\;\;1\leq 2k\leq J\label{odd1}\\
g_{2J+1,2k+1}={J-k\over 2J+1} (3g_{3,1})^{2J-1}\;,\;\;1\leq 2k+1\leq J\label{odd2}\;,
\end{gather}
where $J>1$. It is straightforward to check that (\ref{even1})-(\ref{odd2}) indeed solve all the recursion relations (\ref{recursions})-(\ref{rec3}).

The next natural question is whether the pseudofree action can be written in a simple closed form, rather than as an infinite series. Fortunately, for even couplings $g_{2J,k}$ we don't need to do any work to answer this.
Namely, even couplings given by (\ref{even1}), (\ref{even2}) are equal to those obtained in  \cite{Gabai:2018tmm}, which implies that the even part of the action is the $O(D-2)$ non-linear sigma model 
\be
S_{even}=-{F^2\over 4}\int \mbox{\rm Tr}\,\d_\alpha G\d^\alpha G^{-1}\;,
\ee
where $G$ is an $O(D-2)$ group element in the Cayley parametrization
\be
\label{Cayley}
G={1+{A\over 2F}\over 1-{A\over 2 F}}\;,
\ee
and the axion ``decay constant" $F$ is given by
\be
\label{F}
F={1\over 6 g_{3,1}}\;.
\ee
Given this result, it is natural to expect that the odd part of the action takes the form of the Wess--Zumino (WZ) term \cite{Wess:1971yu,Novikov_multivaluedfunctions,Witten:1983ar},
\be
\label{SWZ}
S_{WZ}={n\over 24\pi}\int d^3\sigma\epsilon^{\alpha\beta\gamma}{\rm Tr }\, G^{-1}\d_\alpha G G^{-1}\d_\beta GG^{-1}\d_\gamma G
\ee
for some rank $n$. To see that this is indeed the case, note first that in the Cayley parametrization (\ref{Cayley}) the WZ term (\ref{SWZ}) turns into
\be
\label{SWZC}
S_{WZ}={n\over 24\pi F^3}\sum_{a,b,c\geq 0}\int d^3\sigma\epsilon^{\alpha\beta\gamma}{\rm Tr}\, \l{A\over 2F}\r^{2a}\d_\alpha A \l{A\over 2F}\r^{2b}\d_\beta A \l{A\over 2F}\r^{2c}\d_\gamma A\;.
\ee
On the other hand the odd part of the action (\ref{gaction}) can be trivially written as a three-dimensional integral of the form
\be
\label{our3d}
S_{odd}=\int d^3\sigma\sum_{J=1}^\infty\sum_{m=1}^{J}g_{2J+1,m}\epsilon^{\alpha\beta\gamma}\d_\gamma\l\Tr\,{\partial_\alpha A A^{m-1}\partial_\beta A A^{2J-m-1}}\r\;.
\ee
It is a matter of a straightforward (even if a bit tedious) calculation to check that for the  values of the coupling constants given in (\ref{odd1}), (\ref{odd2}) the two actions (\ref{SWZC}) and (\ref{our3d}) indeed coincide,
and the rank $n$ is given by 
\be
\label{rank}
n={\pi\over 9g_{3,1}^2} \;.
\ee
Comparing (\ref{SWZ}) and (\ref{rank}) we find that 
the relation
\[
n=4\pi F^2
\]
is satisfied, implying that the pseudofree theory has an identically vanishing $\beta$-function \cite{Witten:1983ar}, which is quite natural. Hence, at integer values of the rank $n$ our search for pseudofree theories has led us in a rather roundabout way to a famous family of conformal theories---WZW models. A posteriori this result is not surprising, given that WZW models are equivalent to free fermionic systems \cite{Witten:1983ar}. This equivalence also supports the expectation that even though our analysis is restricted to tree level diagrams, the resulting theory is pseudofree to all orders in the loop expansion.
It is worth noting also that an observation that tree level four- and five-particle  amplitudes in the WZW model vanish at the conformal point has been made already back in \cite{Figueirido:1988ct}.

\section{Future Directions}
\label{sec:future}
It is intriguing and encouraging that the  logic outlined in the Introduction lead us from QCD strings to a classic rational CFT---the $O(N)$ WZW model. It was envisaged already in \cite{Witten:1983ar} that WZW models can lead to generalizations of conventional critical strings. Indeed, WZW models have been used extensively as a building block for numerous worldsheet  theories since then. An unusual aspect of the construction presented here is that WZW fields transform non-trivially under the Poincar\'e symmetries of flat physical coordinates. Conventionally, additional scalar degrees of freedom on the worldsheet are associated with extra spatial dimensions. Interpreted this way, axionic strings might arise in  a strongly non-factorizable geometry, such that the physical Lorentz symmetry acts both on physical and Kaluza-Klein coordinates.

On the other hand, the appearance of the WZW model suggests that axionic strings may have a natural reformulation in the fermionic language---the WZW model at integer rank is equivalent to a system of free fermions.
In this picture axionic strings start looking very similar to the conventional RNS strings, but to see whether this description is appropriate one needs to understand how to reformulate the  axionic coupling (\ref{SA})  in the fermionic language.

Either way, the next natural step towards understanding axionic strings is to calculate the rank $n$ of the WZW as a function of $D$. At first sight, we have all ingredients to do this. Namely, one may try to combine the relation
(\ref{naiveg31}) with the expression for $Q_A$, proposed in   \cite{Dubovsky:2015zey}, which generalizes (\ref{Qa}) to a general $D$. This allows one to determine $g_{3,1}$, and hence both the WZW decay constant $F$ and rank $n$, as a function of $D$. 

However, it should be clear by now that both (\ref{naiveg31}) and the proposal of  \cite{Dubovsky:2015zey} are too naive. As a result of $\ell_s$ independent self-interactions of the axionic field the relation (\ref{naiveg31}) may  receive an infinite set of loop corrections and the same applies to the expression given in   \cite{Dubovsky:2015zey} at $D
>4$. It definitely looks at this point that this calculation should be done using a natural set of operators present in the WZW model (such as group elements $G$) rather than  perturbatively in $A$. We hope to accomplish this in the near future.

As an interesting byproduct of our analysis, we arrived at a somewhat roundabout construction of the $O(N)$ WZW model. It will be interesting to generalize this analysis and to reconstruct a larger class of CFT's by looking
for  general pseudofree theories starting with a general seed cubic coupling of the form
\[
S_{seed}=\int \epsilon^{\alpha\beta}f_{ijk}\phi^i\d_\alpha\phi^j\d_\beta\phi^k\;.
\]
It looks that if  $f_{ijk}$ are the structure constants of a semisimple Lie algebra, the analysis presented above should go through and will lead to the corresponding WZW model. It is interesting to check whether this exhausts the list of pseudofree models.

{\it Acknowledgements.}  We thank  Victor Gorbenko, Juan Maldacena, Massimo Porrati and Arkady Tseytlin for useful discussions and correspondence. 
This work is supported in part by the NSF CAREER award PHY-1352119.

\appendix
\section{CCWZ Construction for Axionic Strings}
\label{CCWZ}
A straight infinitely long string spontaneously breaks the bulk Poincar\'e group $G=ISO(1,D-1)$ down to $H=ISO(1,1)\times O(D-2)$. In what follows we will employ the static gauge, Greek indices denote the directions along the string and Latin the transverse ones. To find the transformation of the fields we begin by introducing an element of the quotient group $G/H$ in the exponential parametrization 
\be
L=\mbox{exp}\{i\sigma^\alpha P_\alpha+iX^i(\sigma)P_i\}\mbox{exp}\{i\phi^{\alpha i} J_{\alpha i}\}\,.
\label{coselem}
\ee
Under the action of the element $g$ of group $G$ it is transformed as follows
\be
g_1L=L^\prime\,\mbox{exp}\{i\,u_1+i\,v_1\}\,,
\label{coset}
\ee
where the exponent on the r.h.s. is introduced to compensate for a change of the 
representative of the coset and we introduced the notation 
$u_1\equiv u_1^{\gamma\delta}J_{\gamma\delta}$ and $v_1 \equiv v_1^{i j}J_{i j}$. From 
the group action it follows that 
\begin{eqnarray}
g_2L^\prime&=&L^{\prime\prime}\;\mbox{exp}\{i\,u_2+i\,v_2\}\nonumber \\
g_2g_1L&=&L^{\prime\prime}\;\mbox{exp}\{i\,u_2+i\,v_2\}\;
\mbox{exp}\{i\,u_1+i\,v_1\}\nonumber \\
\mbox{exp}\{i\,u_3+i\,v_3\}&=&\mbox{exp}\{i\,u_2+i\,v_2\}\;
\mbox{exp}\{i\,u_1+i\,v_1\}\,,
\end{eqnarray}
which leads to a natural definition of the left G action on the matter fields
\be
\psi^\prime(\sigma^\prime)=D\Big(\mbox{exp}\{i\,u+i\,v\}\Big)\psi(\sigma)\,,
\ee
where $D$ stands for the appropriate representation and in particular 
\be
A^{\prime kl}(\sigma^\prime)=\Big(\mbox{exp}\{i\,v\}\Big)_p^k
\Big(\mbox{exp}\{i\,v\}\Big)_q^l\, A^{pq}(\sigma)\,.
\ee

Thus, to find the transformation of the field $A^{kl}$ we need to find $u$ and $v$,
which can be done by expanding equation (\ref{coset}) to linear order. We define a group element of the non-linearly 
realized boosts 
\be
g=\mbox{exp}\Big\{-i\,\frac{\epsilon}{2} J_{\alpha i} \Big\}\,,
\label{elem}
\ee
for some fixed $J_{\alpha i}$ and use the expressions for the generators and the commutators of the 
Poincar\'e algebra
\begin{eqnarray}
\Big(J_{\rho\sigma}\Big)^\mu_\nu&=&{i}\Big(\eta_{\sigma\nu}\delta_\rho^\mu-
\eta_{\rho\nu}\delta_\sigma^\mu\Big)\nonumber \\
\Big[J_{\mu\nu},P_\rho\Big]&=&-{i}{}\Big(\eta_{\mu\rho}P_\nu-\eta_{\nu\rho}P_\mu\Big)\nonumber \\
\Big[J_{\mu\nu},J_{\rho\sigma}\Big]&=&-{i}{}\Big(\eta_{\mu\rho}J_{\nu\sigma}-
\eta_{\mu\sigma}J_{\nu\rho}-\eta_{\nu\rho}J_{\mu\sigma}
+\eta_{\nu\sigma}J_{\mu\rho}\Big)\,.
\end{eqnarray}
As a result one finds the transformations of the Goldstone fields
\begin{eqnarray}
\delta_{\alpha i}\sigma^\beta&=&\frac{\epsilon}{2}\,X_i\,\delta_\alpha^\beta\nonumber \\
\delta_{\alpha i} X^j&=&-\frac{\epsilon}{2}\,\Big[\delta_i^j\sigma_\alpha
+X_i\d_\alpha X^j\Big]\,,
\label{dxapp}
\end{eqnarray}
and also the recursive relations for $u\,$ and $v$
\be
\Big(1-\frac{i \epsilon}{2}\, J_{\alpha i}\Big)\Big(1+i\phi-\frac{1}{2}\phi^2+\dots\Big)=
\Big(1+i\phi^\prime-\frac{1}{2}\phi^{\prime^2}+\dots\Big)\Big(1+i\,u+i\,v\Big)\,.
\label{phiuv}
\ee

%The remainder of the equation (\ref{coset}) can be used to find the transformation of the 
%auxiliary Goldstone fields $\phi$ as well as 
To the leading order in $\phi$ the solution of (\ref{phiuv}) is 
\begin{eqnarray}
\phi^\prime=\phi-\frac{\epsilon}{2} J_{\alpha i}\,,\quad \mbox{and} \quad
i\,(u+v)=\frac{\epsilon}{4}[J_{\alpha i},\phi]\,.
\label{vij}
\end{eqnarray}
For the action of the group on $A^{k l}$ in particular, we note that explicitly
\be
i\,v^{mn}\,J_{mn}=-i\,\frac{\epsilon}{4}\,\phi^{\beta j}\,J_{ij}\,
\eta_{\alpha\beta}\,.
\label{aux}
\ee
As a final step we need to eliminate the auxiliary fields, which can be done most simply by 
introducing the Maurer-Cartan form
\be
L^{-1}d\,L=d\,\sigma^{\beta}\l ie_{\beta\gamma}P^\gamma+iD_{\beta k}P^k+
iV_\beta^{ij}J_{ij}+iU_\beta^{\gamma\delta}J_{\gamma\delta}+
i\Phi_\beta^{\alpha i}J_{\alpha i}\r\,,
\label{covder}
\ee
and setting the covariant derivative equal to zero 
\be
iD_{\beta k}\,P^k = 0 .
\ee
To find the expression for covariant derivative we substitute expression 
(\ref{coselem}) into (\ref{covder}) 
\be
L^{-1}d\,L=d\,\sigma^\beta\,e^{-i\phi}\Big(iP_\beta+iP_j\d_\beta X^j
+i\d_\beta\phi^{\alpha i}\,J_{\alpha i}\Big)e^{i\phi}\,,
\ee
and expand to leading order in $\phi$
\begin{eqnarray}
e^{-i\phi}\,P_\beta\,e^{i\phi}&\approx&P_\beta-
i\phi^{\alpha i}\Big[J_{\alpha i}\,,P_\beta\Big]=
P_\beta-\phi^{\alpha i}\,P_i\,\eta_{\alpha\beta}\,,\nonumber \\
e^{-i\phi}\,P_j\,e^{i\phi}&\approx&P_j-i\phi^{\alpha i}\Big[J_{\alpha i}\,,P_j\Big]=
P_j+\phi^{\alpha i}\,P_\alpha\,\eta_{ij}\,.
\end{eqnarray}
The covariant derivative is thus
\be
iD_{\beta k}\,P^k = i\,P_i\,\Big(\d_\beta X^i-\phi^{\alpha i}\,\eta_{\alpha\beta}\Big)=0\,,
\ee
hence the auxiliary fields are 
\be
\phi^{\alpha i}\,\eta_{\alpha\beta}=\d_\beta X^i\,.
\ee
Substituting the above expression into (\ref{aux}) finally leads to the leading order transformation 
of $A^{kl}$ field 
\be
A^{\prime kl}(\sigma^\prime)\approx\Big(\mbox{exp}\{\frac{-i\,\epsilon}{4} \partial_{\alpha} X^j J_{i j}\}\Big)_p^k
\Big(\mbox{exp}\{\frac{-i\,\epsilon}{4} \partial_{\alpha} X^j J_{i j}\}\Big)_q^l\, A^{pq}(\sigma)\,,
\ee
or to the first order in $\epsilon$
\be
\delta_{\alpha i} A^{kl}\approx-\frac{\epsilon}{2}\, X_i\,\d_\alpha A^{kl}+
\frac{\epsilon}{4}\,\d_\alpha X^j\l\Big(\eta_{jp}\delta_i^k-\eta_{ip}
\delta_j^k\Big)A^{pl}+\Big(\eta_{jq}\delta_i^l-\eta_{iq}\delta_j^l\Big)A^{kq}\r\,.
\label{daapp}
\ee
As a check, one can observe that the commutators of transformations of $X$ and $A$ satisfy the Poincar\'e algebra, 
\begin{eqnarray}
\Big[\delta_{\beta j},\delta_{\alpha i}\Big]A^{kl}&=&\frac{\epsilon_1\epsilon_2}{4}
\l\eta_{ij}\l\sigma_\beta\d_\alpha A^{kl}-\sigma_\alpha\d_\beta A^{kl}\r+
\eta_{\alpha\beta}\l\l\eta_{ip}\delta_j^k-
\eta_{jp}\delta_i^k\r A^{pl}+\l\eta_{iq}\delta_j^l-\eta_{jq}\delta_i^l\r
A^{kq}\r\r\nonumber \\
\Big[\delta_{\beta j},\delta_{\alpha i}\Big]X^k&=&\frac{\epsilon_1\epsilon_2}{4}
\l\eta_{ij}\l\sigma_\beta\d_\alpha X^k-\sigma_\alpha\d_\beta X^k\r+
\eta_{\alpha\beta}\l \delta_j^kX_i-\delta_i^kX_j\r\r\;.\nonumber
\end{eqnarray}

\bibliographystyle{utphys}
\bibliography{dlrrefs}
\end{document}